\documentclass{article1}
\usepackage{amssymb}
\usepackage{graphicx}
\usepackage{epsfig}
\def\tend{\mathop{\to}}
\def\lim{\mathop{\rm {lim}}}
\usepackage{natbib}
\biboptions{sort&compress}
\begin{document}

\begin{frontmatter}
\title{Proton radius, bound state QED and
the nonlocality of the electromagnetic interaction. }
\author[Kazan]{Renat Kh. Gainutdinov}

\address[Kazan]{Kazan Federal University, Institute of Physics,
Kremlevskaya st.,18, Kazan, Russia}

\begin{abstract}
The result of a recent measurement of the size of the proton [R.
Pohl et al., Nature \textbf{466}, 213] performed on the base of
the muonic hydrogen spectroscopy turned out to be significantly
different, by five standard deviations, from the results derived
from the atomic hydrogen spectroscopy. This large discrepancy
could come from the calculations of the Lamb shift in atomic
hydrogen and muonic hydrogen. Here we show that there is a gap in
the standard bound-state QED that may be the source of the
discrepancy. This gap originates in the fact that within the
framework of this theory the QED corrections are described in
terms of the respective Green functions. The character of the time
evolution of a system which should manifest itself in the general
definition of bound states as stationary states of the system
cannot be described in terms of the Green functions. We present a
consistent way of solving the bound-state problem in QED starting
from the condition of stationarity of the bound states. Formulae
for the energies and the vectors of the states of one-electron
(muon) atoms derived in this way indicate that the standard
bound-state QED does not obey the exact description of the atomic
states and, as a result, the Lamb shift obtained in its framework
should be supplemented by an additional "dynamical" energy shift.
It is shown that in this shift natural nonlocality of the
electromagnetic interaction that in describing the S matrix and
the Green functions is hidden in the renormalization procedure
manifest itself explicitly.
\end{abstract}


\end{frontmatter}

\section*{Introduction}

The development of the theory of quantum electrodynamics was
stimulated by high-precision hydrogen spectroscopy performed by
Lamb and Retherford in 1947 that showed a small splitting between
the $2S_{1/2}$ and $2P_{1/2}$ states known as the Lamb shift.
Despite the theory of QED is the most well-tested, accurate, and
successful theory in physics its mathematical foundation is not
secure because of the still unsolved problem of the ultraviolet
(UV) divergences. The renormalization theory that in QED is used
for removing these divergences actually allows one to pass over
this problem but not to solve them. Richard Feynman said in this
connection: "I think that the renormalization theory is simply a
way to sweep the difficulties of the divergences of
electrodynamics under the rug"\ \cite{nob}. As is well known, in
QED the ultraviolet divergences can be removed from the S matrix
and the Green functions, but cannot be removed from quantities
characterizing the time evolution of quantum systems, since
regularization and renormalization of the scattering matrix leads
to the situation in which divergent terms automatically appear in
the Hamiltonian and hence in the Schr{\"o}dinger equation
\cite{bogol}. From this point of view the results of the recent
muonic hydrogen Lamb shift experiment \cite{Pohl} at PSI that have
led to the proton charge radius significantly different, by five
standard deviations, from the that derived from hydrogen atom
spectroscopy \cite{H_Nier, H_Fis, H_Beau, H_Schw} value seem not
surprising. As Jeff Flowers of Britain's National Physical
Laboratory wrote in Ref. \cite{chink}, if these experimental
discrepancy is confirmed then "high-accuracy work such as that by
Pohl and colleagues, not the high-energy collisions of giant
accelerators, may have seen beyond the standard model of particle
physics".

Here we show that the discrepancy may originate from the fact that
the theory of QED fails in describing the time evolution of a
system. For this reason one is forced to describe the bound states
of QED systems in terms of the S matrix or the Green functions.
However, in general bound states should be determined as
stationary states which evolve in time according to the law:

\begin{equation}\label{stat_law}
\left| {\Psi _n (t)} \right\rangle  = e^{ - iE_n t} \left| {\Psi
_n (0)} \right\rangle .
\end{equation}

If in QED the Hamiltonian, i.e. the operator of the total energy,
were well defined, then from this definition it should immediately
follow that the vectors of the stationary states and their
energies are eigenvectors of the Hamiltonian $ H\left| {\Psi _n }
\right\rangle  = E_n \left| {\Psi _n } \right\rangle $. However,
this is not the case. The time evolution of QED systems remains
uncertain even after renormalization. Actually this demonstrates
the limitation of the existing theory of QED. The problem could be
passed over, provided the bound states of composite systems in QED
could be described in terms of the renormalizable S matrix and the
Green functions. However, the discrepancy between the proton radii
deduced from atomic hydrogen spectroscopy and muonic hydrogen
spectroscopy shows that this may not be the case: the QED
corrections that played an important role in the determination of
these radii had been calculated by using the standard methods of
the bound state QED in which the energy levels of composite
systems are determined by the positions of poles of the respective
Green functions. This makes it necessary to investigate the
problem more precisely. Since locality has been argued to be the
main cause of infinities in QED, it seems natural to resolve this
problem by introducing a nonlocal form factor into the interaction
Hamiltonian density. However, it turned out that such an
introduction of a nonlocal form factor results in the loss of
covariance. The origin of this is the fact that the
Schr{\"o}dinger equation is local in time, and the interaction
Hamiltonian describes an instantaneous interaction. But in
relativistic quantum theory, if we spread the interaction in
space, then we should spread it in time as well. Thus, for the
introduction of nonlocality in the theory to be intrinsically
consistent, a way is needed of solving the evolution problem in
the case when the dynamics in a system is generated by a
nonlocal-in-time interaction. In Ref. \cite{1999} it has been
shown that actually the Schr{\"o}dinger equation is not the only
dynamical equation consistent with the current concept of quantum
physics, and the most general dynamical equation (GDE) consistent
with these principles has been derived. Being equivalent to the
Schr{\"o}dinger equation in the case of instantaneous interaction,
this generalized dynamical equation allows one to generalize the
dynamics to the case where the dynamics is governed by a
nonlocal-in-time interaction.

The GDE is shown to provide a consistent way of solving the
bound-state problem starting from the law~(\ref{stat_law})
determining bound states. In this way formulae are derived that
determine the energies and the vectors of the states of
one-electron (muon) atoms. The values of the atomic energy levels
determined by these formulae differ from the values determined by
the positions of the poles of the Green function of the electron
in the Coulomb field. This difference that will be referred to as
the dynamical shift is the reflection of the fact that there is
not the exact correspondence between the atomic states and the
poles of the Green function. The dynamical shift is shown to get
the contributions from the processes, in which the nonlocal nature
of the electromagnetic interaction that in describing the S matrix
and the Green function is hidden in the renormalization procedure
manifests itself explicitly.

\section{Quantum mechanical rules and the generalized dynamical equation}

Paraphrasing Steven Weinberg \cite{Wien} the recent situation in
QED may be characterized as follows. If it turned out some QED
system could not be described by the theory of QED, it would be a
sensation. However, if it turned out that the system did not obey
the rules of quantum mechanics and relativity, it would be a
cataclysm. Of course, the question is raised, in this connection,
what are the basic quantum mechanical rules that must be satisfied
in any theory. In the Feynman`s book \cite{Feyn_book} where a
minimal set of physical principles that must be satisfied in any
theory of fields and particles is analyzed, the only quantum
mechanical principle included in this set is the principle of the
superposition of probability amplitudes. This principle formulated
as the result of analysis of the phenomenon of quantum
interference \cite{Feyn_1948} reads as follows.

The probability of an event is the absolute square of a complex
number called the probability amplitude. The probability amplitude
of an event which can happen in several different ways is a sum of
the probability amplitudes for each of these ways.

In the Feynman formulation of quantum theory \cite{Feyn_1948, FH}
this principle is used as a basic postulate. This postulate
provides the general rule prescribing how to calculate
probabilities in quantum theory, and can be used in different ways
depending on the choice of the class of alternative ways in which
events can happen. In the Feynman formalism the processes
associated with completely specified paths of particles in
space-time are used as such alternatives. The contribution from a
single path is postulated to be an exponential whose (imaginary)
phase is the classical action (in units of $\hbar$) to the path in
question. Thus in the case of such a choice of the class of
alternatives the contribution of each of the alternative way must
be specified from the every beginning. However, very surprisingly
it has turned out \cite{1999} that there is a much more general
class of alternatives whose contributions need not to be
postulated: it is enough to know a priori only the contributions
from the alternative processes associated with a fundamental
interaction while the contributions from other alternatives are
determined by the requirement of the conservation of probability.
This class of alternatives consists of the processes with
completely specified instants of the beginning and end of
interaction in a quantum system. With such a class of
alternatives, the superposition principle allows one to represent
the probability amplitude of finding a quantum system in the state
$ \left| {\Psi _2 } \right\rangle $ at time $t$, if at time $t_0$
it was in the state $ \left| {\Psi _1 } \right\rangle $, in the
form \cite{1999}

\begin{equation}\label{evo}
 \langle \Psi _2 | U(t,t_0 )|\Psi _1 \rangle = \langle \Psi _2 | \Psi _1 \rangle +
\int\limits_{t_0 }^t dt_2 \int\limits_{t_0 }^{t_2 } dt_1 \langle
\Psi _2 |\tilde {S}(t_2 ,t_1 )| \Psi _1 \rangle.
\end{equation}
Here $\langle \Psi _2 |\tilde {S}(t_2 ,t_1 )| \Psi _1 \rangle$ is
the probability amplitude that the interaction in the system
begins in the time interval $(t_1, t_1 + dt_1)$ and ends in the
time interval $(t_2, t_2 + dt_2)$, and after the end of the
interaction the system will be found in the state $ \left| {\Psi
_2 } \right\rangle $ (we use the interaction picture), if before
the beginning of the interaction the system was in the state $
\left| {\Psi _1 } \right\rangle $. By using the operator
formalism, one can represent amplitudes $<\Psi_2|
U(t,t_0)|\Psi_1>$ by the matrix elements of the unitary evolution
operator $ U(t,t_0) $ in the interaction picture. The operator $
\tilde {S}(t_2 ,t_1 ) $ defined in the same way represents the
contribution to the evolution operator from the process in which
the interaction in the system begins at time $t_1$ and ends at
time $t_2$. As has been shown in Ref. \cite{1999}, for the
evolution operator in the form~(\ref{evo}) to be unitary for any
$t$ and $t_0$ the operator $ \tilde {S}(t_2 ,t_1 ) $ must satisfy
the equation
\begin{equation}\label{main}
(t_2 - t_1 )\tilde {S}(t_2 ,t_1 ) = \int\limits_{t_1 }^{t_2 }
{dt_4 \int\limits_{t_1 }^{t_4 } {dt_3 (t_4 - t_3 )} } \tilde
{S}(t_2 ,t_4 )\tilde {S}(t_3 ,t_1 ).
\end{equation}
A remarkable feature of this equation is that it works as a
recurrence relation and allows one to obtain the operators $
\tilde {S}(t_2 ,t_1 ) $ for any $t_1$ and $t_2$, if $\tilde S(t'_2
,t'_1 )$ corresponding to infinitesimal duration times $\tau  =
t'_2  - t'_1$ of interaction are known. It is natural to assume
that most of the contribution to the evolution operator in the
limit $t_2 \to t_1$ comes from the processes associated with the
fundamental interaction in the system under study. Denoting this
contribution by $H_{int}(t_2,t_1)$ we can write
\begin{equation}\label{Stend}
\tilde {S}(t_2 ,t_1)\tend\limits_{t_2 \to t_1} H_{int} (t_2 ,t_1)
+ O(\tau^\epsilon),
\end{equation}
where $\tau =  t_2 - t_1$.  The parameter $\epsilon$ is determined
by demanding that $H_{int}(t_2,t_1)$ called the generalized
interaction operator must be so close to the solution of Eq.
(\ref{main}) in the limit $t_2 \to t_1$ that this equation has a
unique solution having the behavior (\ref{Stend}) near the point
$t_2 = t_1$. Actually, Eqs.~(\ref{evo}) and~(\ref{main}) represent
the quantum mechanical rules that must be obtained by any physical
theory. Equation~(\ref{evo}) is the representation of the
probability amplitude $<\Psi_2| U(t,t_0)|\Psi_1>$ as a sum of the
contributions from all alternative ways in which the event can
happen for the chosen class of alternatives, and Eq.~(\ref{main})
is simply the unitarity condition in terms of these contributions.
The wonderful feature of this rule is that it directly leads to
the dynamical equation. If $H_{int}(t_2,t_1)$  is specified, the
"unitarity condition"\~(\ref{main}) allows one to determine $
\tilde {S}(t_2 ,t_1 ) $ for any $t_1$ and $t_2$ and hence to
construct the evolution operator. Thus, being supplemented by the
boundary condition~(\ref{Stend}), the relationship~(\ref{main})
becomes the equation of motion for states of a quantum system. It
is important that this equation is universal, and the specific
features of a theory describing the dynamics of quantum systems
manifest themselves only in the boundary condition~(\ref{Stend})
and in the Hilbert space with which the theory is dealing. In the
case when the quantum field theory is required for the description
of physical processes the Hilbert space must be chosen in the form
of the Fock space, and correspondingly the interaction operator
$H_{int}(t_2,t_1)$ should be constructed in terms of the field
operators. In the case when the interaction operator is of the
form
\begin{equation}\label{delta}
H_{int} (t_2 ,t_1 ) = - 2\pi i\delta (t_2 - t_1 )H_I (t_1),
\end{equation}
i.e., the fundamental interaction is instantaneous, the
generalized dynamical equation~(\ref{main}) turns out to be
equivalent \cite{1999} to the Schr{\"o}dinger equation with the
interaction operator $ H_I (t_1)$. A quantum field theory must
also obey the relativity. This condition is satisfied in the case
when the interaction Hamiltonian density ${\cal H}(x)$ related to
the interaction Hamiltonian $ H_I (t_1)$ by the equation $ H_I (t)
= \int {d^3 x} {\cal{H}}_I(t = 0,{\bf{x}}) $ is local, i.e. the
interaction is local both in time and in space. But this locality
leads to the UV divergences. At the same time the interaction
operator of the form~(\ref{delta}) is the specific case of
operators allowed by Eq.~(\ref{Stend}). In general the interaction
operators describe the interaction spread both in space and in
time. In this case the dynamics of a quantum system is determined
by the behaviour of the interaction operator in the
Schr{\"o}dinger picture, ${\cal H}^{(s)}_{int}(x)$ , in the limit
of infinitesimal duration times $\tau$ of the interaction (in the
local case the dynamics is determined by the interaction operator
specified at the point $ \tau  = 0 $). Thus there is no finite
scale of nonlocality in the case of such nonlocal interaction! In
order to obey these rules the operator describing the time
evolution in QED should be of the form~(\ref{evo}) with $ \tilde
{S}(t_2 ,t_1 ) $ satisfying Eq.~(\ref{main}), and hence the
fundamental QED interaction should be described by one of the
nonlocal interaction operators allowed by Eq.~(\ref{Stend}) and
obeying the rule of relativity. Thus the problem of the UV
divergences is reduced to the problem of finding the form of this
operator.

Formally Eq.~(\ref{evo}) involves processes associated with
vacuum-vacuum transitions. Correspondingly the matrix element $
\left\langle {\Psi _2 } \right|U(t,0)\left| {\Psi _1 }
\right\rangle$ of the evolution operator defined by
Eq.~(\ref{evo}) is the product of the probability amplitude $
\left\langle {\Psi _2 } \right|U_{ph}(t,0)\left| {\Psi _1 }
\right\rangle$ of the physical event under study and the
probability amplitude of the vacuum-vacuum transition:
\[ \left\langle {\Psi _2 } \right|U(t,0)\left| {\Psi _1 }
\right\rangle  = \left\langle {\Psi _2 } \right|U_{ph} (t,0)\left|
{\Psi _1 } \right\rangle \left\langle 0 \right|U(t,0)\left| 0
\right\rangle. \] This one-loop contribution is proportional to
the space-volume $V$ because of the translational invariance. For
the same reason the contribution from the two-loop vacuum process
is proportional to $V^2$ and so on. The amplitude $ \left\langle
{\Psi _2 } \right|U_{ph}(t,0)\left| {\Psi _1 } \right\rangle$, of
course, is a sum of contributions from the process with the
specified moments of the beginning and end of the interaction

\[
\left\langle {\Psi _2 } \right|U_{ph} (t,0)\left| {\Psi _1 }
\right\rangle  = \left\langle {\Psi _2 } \right.\left| {\Psi _1 }
\right\rangle  + \int\limits_{t_0 }^t {dt_2 } \int\limits_{t_0
}^{t_2 } {dt_1 } \left\langle {\Psi _2 } \right|\tilde S_{ph} (t_2
,t_1 )\left| {\Psi _1 } \right\rangle
\]

However, in this case the processes of the interaction associated
with the vacuum-vacuum transitions must not be taken into account.
Correspondingly, $ \tilde {S}_{ph}(t_2 ,t_1 ) $ must satisfy  Eq.
~(\ref{main}), from the right-hand part of which one has to remove
the terms proportional to $V^n$, i.e., the terms associated with
the processes which involve the vacuum-vacuum transitions:
\[
(t_2  - t_1 )\tilde S_{ph} (t_2 ,t_1 ) = \int\limits_{t_1 }^{t_2 }
{dt_4 \int\limits_{t{}_1}^{t_4 } {dt_3 (t_4  - t_3 )\tilde S_{ph}
(t_2 ,t_4 )} } \tilde S_{ph} (t_3 ,t_1 ) - c.t.
\]
where \textit{c.t.} stands for counter terms proportional to
$V^n$.

In the further discussion we will use the general notation $
\left\langle {\Psi _2 } \right|U(t,0)\left| {\Psi _1 }
\right\rangle$ and $ \left\langle {\Psi _2 } \right|\tilde S(t_2
,t_1 \left| {\Psi _1 } \right\rangle$ for describing the
"physical"\ processes and the GDE in the form~(\ref{main}) keeping
in mind that the contribution from the above terms in this
equation must not be taken into account.

It is extremely important that, being the representation of the
general quantum mechanical rules, Eqs.~(\ref{evo}) and
~(\ref{main}) allow one to obtain detailed information about
physical processes without specifying the interaction operator.
Such information can be regarded as a direct consequence of the
first principles. Let us now investigate, in this way, the
bound-state QED.

Expression~(\ref{evo}) for $ \left\langle {\Psi _2 }
\right|U(t,0)\left| {\Psi _1 } \right\rangle$ in the
Schr{\"o}dinger picture can be rewritten in the form
\begin{equation}\label{u_g}
\left\langle {\Psi _2 } \right|U_S (t,0)\left| {\Psi _1 }
\right\rangle  = \frac{1}{{2\pi }}\int\limits_{ - \infty }^\infty
{dE} e^{ - iEt} \left\langle {\Psi _2 } \right|G(E + i0)\left|
{\Psi _1 } \right\rangle ,
\end{equation}
where
\begin{equation}\label{g_t}
G(z) = G_0 (z) + G_0 (z)T(z)G_0 (z)
\end{equation}
with $ G_0(z) = (z - H_0)^{-1}$ and
\begin{equation}\label{t_s-}
T(z) = i\int\limits_0^\infty  {d(t_2  - t_1 )\exp [iz(t_2  - t_1
)]} \exp [ - iH_0 t_2 ]\tilde S(t_2 ,t_1 )\exp [iH_0 t_1 ]
\end{equation}
In terms of the T operator defined by Eq.~(\ref{t_s-}) the
generalized dynamical equation~(\ref{main}) can be rewritten in
the form \cite{1999}
\begin{equation}\label{dt_dz}
\frac{{d < \psi _2 |T(z)|\psi _1  > }}{{dz}} =  - \sum\limits_n
{\frac{{ < \psi _2 |T(z)|n >  < n|T(z)|\psi _1  > }}{{(z - E_n )^2
}}}
\end{equation}
where $n$ stands for the entire set of discrete and continuous
variables that characterize the system in full, and $|n\rangle$
are the eigenvectors of $ H_0 $. Correspondingly, the boundary
condition~(\ref{Stend}) takes the form
\begin{equation}\label{gran_tz_bz}
 < \psi _2 |T(z)|\psi _1  > \tend \limits_{|z| \tend \infty} < \psi _2 |B(z)|\psi _1
 >,
\end{equation}

\begin{equation}\label{bz_ht}
B(z) = i \int_0^{\infty} d\tau exp(iz \tau) H^{(s)}_{int}(\tau),
\end{equation}
where
\begin{equation}\label{h_hint}
H^{(s)}_{int}(\tau) = exp(-iH_0t_2) H_{int}(t_2,t_1) exp(iH_0t_1)
\end{equation}
is the interaction operator in the Schr{\"o}dinger picture.
Equation~(\ref{u_g}) coincides in its form with the standard
expression that relates the evolution operator to the Green
operator. In the canonical formalism the Green operator is defined
as the resolvent of the total Hamiltonian $H$:
\begin{equation}\label{gz}
G(z) = (z - H)^{-1}
\end{equation}
and Eq.~(\ref{g_t}) is the definition of the T-matrix. On the
other hand, Eq.~(\ref{u_g}) is simply one of the forms of
Eq.~(\ref{evo}) representing the superposition principle, where
the Green operator $G(z)$ which in general can be defined as the
inverse Fourier transform of the evolution operator, given by Eq.
~(\ref{u_g}), can be represented in the form~(\ref{g_t}). The
Green operator defined in this way coincides with $G(z)$ defined
by Eq.~(\ref{gz}) only in the case when the interaction in the
system is instantaneous, i.e., in the case when the GDE
~(\ref{main}) is reduced to the Schr{\"o}dinger equation. This
definition of the Green operator is valid even in the case when
the total Hamiltonian $H$  makes no sense. This is very important
because the Green operator $G(z)$ characterizes the evolution of a
system, and, in particular, the positions of its poles determine
the energies of stationary states, and this makes room for solving
the bound-state problem in QED in a consistent way.

\section{The bound-state QED and the Green operator}

In solving the bound state problem in QED it is convenient to
include the Coulomb field into the free Hamiltonian from the every
beginning. This leads us to the Furry picture in which the
eigenstates $|n\rangle$ of the Dirac-Coulomb Hamiltonian $H^D_0$
($ H_0^D \left| n \right\rangle  = E_n^{(0)} \left| n
\right\rangle $) are used as "free" states. The one-electron
states $\left| {\Psi _n^{(0)} } \right\rangle$ corresponding to
the discrete spectrum of energy are just the "bare" states of a
one-electron atom, i.e. the atomic states, when the atom does not
interact with the vacuum. The real atomic states are results of
dressing the bare states $\left| {\Psi _n^{(0)} } \right\rangle$
by this interaction. As we show below, the GDE provides a new very
effective way of solving this problem.

In QED the operator $T(z)$ describes not only the interaction
between particles but also their self-action. This problem can be
overcome by reduction, which amounts to the propagator $G_0(z)$
describing the evolution of free particles being replaced by the
propagator $ \tilde G_0 (z)$ describing the evolution of particles
interacting with the vacuum and, accordingly $T(z)$ being replaced
by $T(z)$, which describes particle interaction proper. Redefined
in this way the free Green operator $G_0(z)$ describes the
evolution of the system in the case when the particles move freely
or interact only with the vacuum. In this way the Green operator
can be rewritten in the form
\begin{equation}\label{g_m}
G(z) = \widetilde G_0 (z) + \widetilde G_0 (z)M(z)\widetilde G_0
(z)
\end{equation}
From Eq.~(\ref{g_m}) it follows that the "free" Green operator $
\tilde G_0 (z)$ should be of the form
\begin{equation}\label{gz-}
\widetilde G_0 (z) = \sum\limits_m {\frac{{\left| m \right\rangle
\left\langle m \right|}}{{z - E_m ^{(0)}  - C_m (z)}}}
\end{equation}
and the GDE can be rewritten in the terms of $ C_m (z)$ and the
operator $ M (z)$:
\begin{equation}\label{dc_dz}
\left\langle {{m'}}
 \mathrel{\left | {\vphantom {{m'} m}}
 \right. \kern-\nulldelimiterspace}
 {m} \right\rangle \frac{{dC_m (z)}}{{dz}} =  -  < m'|P_m M(z)\left( {\tilde G_0 (z)} \right)^2 M(z)|m
 >,
\end{equation}

\begin{equation}\label{dm_dz}
\begin{array}{l}
 \frac{{d < m'|M(z)|m > }}{{dz}} =  -  < m'|P_m^ \bot  M(z)\left( {\tilde G_0 (z)} \right)^2 M(z)|m >  -  \\
  -  < m'|P_m^ \bot  \frac{{dC_{m'} (z)}}{{dz}}\tilde G_0 (z)M(z)|m >  -  < m'|P_m^ \bot  M(z)\tilde G_0 (z)\frac{{dC_m (z)}}{{dz}}|m > , \\
 \end{array}
\end{equation}
where $ P_n $ is the projection operator on the state $\left|
{\Psi _n^{(0)} } \right\rangle$ and $P_n  + P_n^ \bot   = {\bf{1}}
$. The corresponding boundary conditions are as follows
\begin{equation}\label{m_pb}
 < m'|M(z)|m > \tend \limits_{|z| \tend \infty} < m'|P_m^ \bot  B(z)|m >,
\end{equation}

\begin{equation}\label{c_pb}
 < m'|m > C_m (z)\tend \limits_{|z| \tend \infty} < m'|P_m^{} B(z)|m >,
\end{equation}

Now we can obtain the vector of the corresponding atomic state.
The stationarity of the bound state of an electron in the Coulomb
field $ \left| {\Psi _n^{} (t)} \right\rangle$ with the energy $
E_n$ must manifest itself in the fact that the Green operator $ G
(z)$ has a pole at $z = E_n$ and a residue being the projection
operator on this state:
\begin{equation}\label{gz2}
G(z) = \frac{{\left| {\Psi _n^{} } \right\rangle \left\langle
{\Psi _n^{} } \right|}}{{z - E_n ^{} }} + {\rm O}(1), \qquad z \to
E_n ^{}
\end{equation}
The positions of the poles of the Green operator of the form
~(\ref{g_m}) are determined by the positions of the poles of the
operator $ \tilde G_0 (z)$ that in turn are determined by the
equation
\begin{equation}\label{pole_eq2}
E_n  - E_n^{(0)}  - C_n (E_n ) = 0.
\end{equation}
where $n$ corresponds to the discrete spectrum. For the positions
of these poles to determine the atomic states $ \left| {\Psi _ n}
\right\rangle$, their residues should be the projections on these
states, $ \left| {\Psi _n } \right\rangle \left\langle {\Psi _n }
\right|$. In order to find these residues, one has to analyze all
of the contributions to the Green operator having such a pole.
Besides the operator $ \tilde G_0 (z)$, this singularity obviously
takes place in
$$ P_n^{} \widetilde G_0 (z)M(z) \equiv P_n^{}
\widetilde G_0 (z)M(z)P_n^ \bot$$ and
$$ M(z)\widetilde G_0(z)P_n^{} \equiv P_n^ \bot  \widetilde G_0 (z)M(z)P_n^{}.$$
Note that $ < \Psi _n^{(0)} |M(z)|\Psi  >  =  < \Psi _n^{(0)}
|M(z)P_n^ \bot  |\Psi >$ and $ < \Psi |M(z)|\Psi _n^{(0)}  > =$
$<\Psi |P_n^ \bot  M(z)|\Psi _n^{(0)} > $ by the definition. The
singularity in the part of the Green operator given by $ P_n^ \bot
\widetilde G_0 (z)M(z)\widetilde G_0 (z)P_n^ \bot$ can arise only
because of the singularity of the operator $ P_n^ \bot M(z)P_n^
\bot$. It is very important that Eqs.~(\ref{dc_dz}) and
~(\ref{dm_dz}) allow to determine the behavior of the operator
$T(z)$ in the vicinity of the singular point $z = E_n$ without
specifying the interaction operator, and this behavior is as
follows
\begin{equation}\label{mmmmm}
\begin{array}{l}
 M(z) = \frac{{C_0 }}{{(z - E_n )}}\left\{ {M(E_n )|\Psi _n^{(0)}  >  < \Psi _n^{(0)} |M(E_n ) + P_n^ \bot  M(z)P_n^{}  + P_n^{} M(z)P_n^ \bot  } \right\} +  \\
  + {\rm O}(1),\qquad z \to E_n  \\
 \end{array}
\end{equation}
where $ C_0  = \left( {1 - \frac{{dC_m (z)}}{{dz}}_{\left| {z =
E_n } \right.} } \right)^{ - {\raise0.7ex\hbox{$1$}
\!\mathord{\left/
 {\vphantom {1 2}}\right.\kern-\nulldelimiterspace}
\!\lower0.7ex\hbox{$2$}}}$

The insertion of this expression in Eq.~(\ref{g_m}) yields
\begin{equation}\label{gz3}
G(z) = \frac{{\left| {\Psi'_n } \right\rangle \left\langle {\Psi
_n^{} } \right|}}{{z - E_n ^{} }} + {\rm O}(1),\qquad z \to E_n
^{}
\end{equation}
with
\begin{equation}\label{psi}
\left| {\Psi _n^{} } \right\rangle  = \overline C _0 \left(
{\left| {\Psi _n^{(0)} } \right\rangle  + P_n^ \bot \widetilde
G_0^ +  (E_n )M^ +  (E_n )\left| {\Psi _n^{(0)} } \right\rangle }
\right)
\end{equation}

\begin{equation}\label{psi_s}
\left| {\Psi '_n } \right\rangle  = C_0 \left( {\left| {\Psi
_n^{(0)} } \right\rangle  + P_n^{\bot} \widetilde G_0 (E_n)
M(E_n)\left| {\Psi _n^{(0)} } \right\rangle } \right)
\end{equation}
At $z = E_1$  ($n = 1 $ corresponds to the ground state of the
atom) the operators $ \tilde G_0 (z)$ and $M(z)$ are Hermitian
and, as a result, $\left| {\Psi '_1 } \right\rangle  = \left|
{\Psi _1 } \right\rangle$. From this and the fact that the vector
$ \left| {\Psi _ 1} \right\rangle$ is normalized $\left\langle
{{\Psi '}} \mathrel{\left | {\vphantom {{\Psi '} {\Psi _1 }}}
\right. \kern-\nulldelimiterspace} {{\Psi _1 }} \right\rangle  =
1$, and Eq.~(\ref{gz3}), it follows that the ground state $\left|
{\Psi _1 (t)}\right\rangle$ is stationary: $\left| {\Psi _1 (t)}
\right\rangle  = e^{ - iE_1 t} \left| {\Psi _1 (0)}
\right\rangle$. The difference $\delta E_1^{(0)}  = C(E_1)$
between the energy $E_1$ of this state and $E^{(0)}_1$ is caused
by the interaction of the atom with the vacuum, and is just the
Lamb shift of the energy level of the ground state. The vector $
\left| {\Psi _ 1} \right\rangle$ describes the ground state of the
atom. This is not a one-electron state even in the sense of the
Dirac equation. Actually in this case we deal with the state of
the atom surrounded by a cloud of virtual particles. The second
term in expression~(\ref{psi}) for the state $ \left| {\Psi _ 1}
\right\rangle$ just describes the probability to find the virtual
particles such as, for example, photons and electron-positron
pairs in a measurement, if before the experiment the atomic system
was in the state $ \left| {\Psi _ 1} \right\rangle$.

The exited states $\left| {\Psi _n (t)}\right\rangle$ ($n = 2, 3,
…$) can be regarded as quasistationary states because their
energies determined by Eq.~(\ref{pole_eq2}) are complex $E_n  =
E_n^{(0)}  + C_n (E_n ) \equiv E_n^{(0)}  + \delta E_n^{}  -
i\Gamma _n /2$. This results in the fact that the states $ \left|
{\Psi '_n } \right\rangle$ and $ \left| {\Psi_n } \right\rangle$
given by Eqs.~(\ref{psi}) and~(\ref{psi_s}) respectively do not
coincide exactly.

Our analysis of the unstable atomic states can be extended to the
general description in terms of the energy distribution
\cite{Krylov}. However, in this work we will restrict ourselves to
the approximation that is proven to be applicable for solving the
most of the problems in quasistationary atomic physics. In this
approximation the difference between the vectors $ \left| {\Psi
'_n } \right\rangle$ and $ \left| {\Psi_n } \right\rangle$ can be
neglected.

Formulae~(\ref{pole_eq2}) and~(\ref{psi}) determining the energies
$E_n$ and the vectors $ \left| {\Psi_n } \right\rangle$ f the atom
have been derived as an inevitable consequence of Eqs.
~(\ref{dc_dz}) and~(\ref{dm_dz}), which in turn represent the
basic quantum mechanical rules, without resorting to any specific
information about the character of the electromagnetic
interaction. In order to use these formulae for the calculations
one has to obtain, at first, the operator $M(z)$ and the function
$C_n(z)$ by solving Eqs.~(\ref{dc_dz}) and~(\ref{dm_dz}). Only at
this stage of our analysis we have to specify the interaction
operator. The interaction operator $B(z)$ that describes the local
electromagnetic interaction is of the form
\begin{equation}\label{b_hi}
B(z) = H_I = \int {d^3 x} {\cal{H}}_I(0,{\bf{x}}) = e\int {d^3 x}
:A_\mu  (0,{\bf{x}})\bar \Psi (0,{\bf{x}})\gamma ^\mu  \Psi
(0,{\bf{x}}):
\end{equation}
where ${\cal{H}}_I(x) $ is the interaction Hamiltonian density,
$\Psi (t,{\bf{x}})$ is the Dirac field in the Furry picture and
$A^\mu(x)$ is the electromagnetic field. The value of the coupling
constant of the electromagnetic interaction allows to solve
Eqs.~(\ref{dc_dz}) and~(\ref{dm_dz}) perturbatively. In the first
order of this solution we have
\begin{equation}\label{m_hi}
M^{(1)} (z) = H_I, \qquad \tilde G_0^{(1)} (z) = G_0 (z)
\end{equation}
In the next order, for $C_n(z)$ we have the equation
\begin{equation}\label{dc2_dz}
\frac{{dC_n^{(2)} (z)}}{{dz}} =  -  < \Psi _n^{(0)} |H_I \left(
{G_0 (z)} \right)^2 H_I |\Psi _n^{(0)}  >
\end{equation}
The form of the interaction operator~(\ref{b_hi}) implies that
$C_n(z)$ tends to zero as $|z| \to \infty$. Solving Eq.
~(\ref{dc2_dz}) with this boundary condition yields
\begin{equation}\label{c2}
\begin{array}{l}
 C_n^{(2)} (E) = \int\limits_0^\infty  {e^{iE_n (t_2  - t_1 )} } C_n (t_2  - t_1 )d(t_2  - t_1 ) = \int\limits_0^\infty  {d(t_2  - t_1 )} \int {d^3 {\bf{p}}_1 } \int {d^3 {\bf{p}}_2 } \bar \Psi _n ({\bf{p}}_2 ) \times  \\
  \times \left[ {\Sigma _A ({\bf{p}}_2 ,t_2 ;{\bf{p}}_1 ,t_1 )e^{iE_n (t_2  - t_1 )}  + \Sigma _A ({\bf{p}}_2 ,t_1 ;{\bf{p}}_1 ,t_2 )e^{i\left\{ {(E - 2E_n^{(0)} )(t_2  - t_1 )} \right\}} } \right]\Psi _n ({\bf{p}}_1 ) \\
 \end{array}
\end{equation}
where
\begin{equation}\label{sigma}
\Sigma _A ({\bf{p}}_2 ,t_2 ;{\bf{p}}_1 ,t_1 ) = \int {d^4 x_1 }
\int {d^4 x_2 } \Sigma _A (x_2 ,x_1 )e^{i{\bf{p}}_2 {\bf{x}}_2 }
e^{ - i{\bf{p}}_1 {\bf{x}}_1 }
\end{equation}
with $\Sigma _A (x_2 ,x_1 )$ being the ordinary one-loop
self-energy operator of the electron. The leading order $C_n(E)$
obtained in this way can then be used for obtaining the leading
order energy shift. For this one has to solve Eq.~(\ref{pole_eq2})
which in this case is reduced to the equation
\begin{equation}\label{pole_eq3}
E - E_n^{(0)}  - C_n^{(2)} (E + i0) = 0
\end{equation}
Neglecting the dependence of $C^{(0)}_n(z)$ on $z$ in the vicinity
of the point $z = E^{(0)}_n$, from this equation for the leading
order self energy correction $\delta E_n  = E - E_n^{(0)}$, we get
\begin{equation}\label{delta_E}
\delta ^{(0)} E_n  = C_n^{(2)} (E_n^{(0)}  + i0) = \int {d^3
{\bf{p}}_1 } \int {d^3 {\bf{p}}_2 } \bar \Psi _n ({\bf{p}}_2
)\Sigma _A ({\bf{p}}_2 ,{\bf{p}}_1 ,E_n^{(0)} )\Psi _n ({\bf{p}}_1
)
\end{equation}
Here we have taken into account that $\Sigma _A ({\bf{p}}_2 ,t_2
;{\bf{p}}_1 ,t_1 )$ depends on the difference $(t_2-t_1)$ only. Of
course, the one-loop self-energy operator $\Sigma _A ({\bf{p}}_2
,{\bf{p}}_1 ,E_n^{(0)} )$ makes no mathematical sense because of
the UV divergences. But the renormalization theory is applicable
in this case, and the problem is solved by replacing $\Sigma _A
({\bf{p}}_2 ,{\bf{p}}_1 ,E_n^{(0)} )$ in Eq.~(\ref{delta_E}) by
its renormalized value $\Sigma^{(R)} _A ({\bf{p}}_2 ,{\bf{p}}_1
,E_n^{(0)} )$. Eq.~(\ref{delta_E}) coincides with the well known
expression for the one-loop energy shift in hydrogen that in the
standard bound state QED is derived, for example, from the
solution of the effective Dirac equation with the mass operator
\cite{Akh}
\begin{equation}\label{Dirac}
H_0^{\rm{D}} \Psi (x) + i\int {\gamma _0 \Sigma _A (x,x',E)\Psi
(x')dx'}  = E\Psi (x)
\end{equation}
where $H_0^{\rm{D}}$ is the Dirac Hamiltonian without the mass
operator but with the external field. In fact, in the leading
order of the expansion of the solution of Eq.~(\ref{Dirac}) in
powers of $\alpha$ we can put $\Psi (x) = \Psi _0 (x)$, and
$\Sigma _A ({\bf{x}},{\bf{x}}',E) = \Sigma _A
({\bf{x}},{\bf{x}}',E_0 )$ in this equation, and in this way we
arrive at the expression~(\ref{delta_E}) for the self-energy
correction. However,~(\ref{delta_E}) is not an exact one-loop
energy shift, because $E = E_n^{(0)} + C_n^{(0)} (E_n )$ is only
an approximate solution of Eq.~(\ref{pole_eq2}). At the next order
of the iterative solution of this equation we get $z = E_n^{(0)} +
C_n^{(2)} (E'_n)$. By using Eq.~(\ref{c2}) and the representation
\begin{equation}\label{sigma2}
\Sigma _A ({\bf{x}}_2 ,{\bf{x}}_1 ;t_2 ,t_1 ) = \frac{1}{{2\pi
}}\int {dEe^{ - iE(t_2  - t_1 )} } \Sigma _A ({\bf{x}}_2
,{\bf{x}}_1 ;E)
\end{equation}
at this order for the one-loop correction, we get
\begin{equation}\label{delta_E2}
\delta E^{(2)}_n  = \int {d^3 {\bf{p}}_1 } \int {d^3 {\bf{p}}_2 }
\bar \Psi _n ({\bf{p}}_2 )\Sigma _A ({\bf{p}}_2 ,{\bf{p}}_1
,E'_n)\Psi _n ({\bf{p}}_1 ) + \delta ^{\rm{D}} E_n ,
\end{equation}
where
\begin{equation}\label{delta_Ed}
\delta ^{\rm{D}} E_n  = \int {d^3 {\bf{p}}_1 } \int {d^3
{\bf{p}}_2 } \bar \Psi _n ({\bf{p}}_2 )\Sigma ^{(n)} ({\bf{p}}_2
,{\bf{p}}_1 )\Psi _n ({\bf{p}}_1 )
\end{equation}
with
\begin{equation}\label{sigma_n}
\Sigma ^{(n)} ({\bf{p}}_2 ,{\bf{p}}_1 ) =  - \frac{{\delta
_n^{(0)} }}{{2\pi i}}\int\limits_{}^{} {dE} \frac{{\Sigma _A
({\bf{p}}_2 ,{\bf{p}}_1 ,E)}}{{\left( {E_n^{(0)}  - E + i0}
\right)^2 }}
\end{equation}

The first term on the right-hand part of Eq.~(\ref{delta_E2}) is
the energy shift, which is determined by Eq.~(\ref{delta_E}), when
in this equation we put $\Sigma _A ({\bf{x}},{\bf{x}}',E) = \Sigma
_A ({\bf{x}},{\bf{x}}',E_0 )$. Thus, there is a discrepancy
between the predictions of the standard bound state QED and our
approach.

The discrepancy is the manifestation of the fact that actually
there is no one-to-one correspondence between the poles of the
Green operator and the poles of the respective Green functions:
Eq. (34) determines the poles of the Green function of the
electron in the Coulomb field. To explain this point note that the
electronic Green function in the Furry picture is determined as
\begin{equation}\label{add1}
G_{ik} (x,x_0 ) = i\left\langle 0 \right|T\Psi _i^{(e)} (x)\bar
\Psi _k^{(e)} (x_0 )S\left| 0 \right\rangle
\end{equation}
where $S$ is the scattering matrix and $\Psi _i^{(e)} (x)$ and
$\Psi _k^{(e)} (x_0 )$ are the electronic field operators in the
Furry picture, and $i,j$ are spin indexes, in the one-loop
approximation under study, the S matrix is given by the first two
terms in the Feynman-Dyson expansion, and, as a result, the Green
function takes the form
\begin{equation}\label{add2}
\begin{array}{l}
 G_{ik} (x,x_0 ) = G_{ik}^{(2)} (t,{\bf{x}};t_0 ,{\bf{x}}_0 ) = i\left\langle 0 \right|T\Psi _i^{} (t,{\bf{x}})\bar \Psi _k^{} (t_0 ,{\bf{x}}_0 )\left| 0 \right\rangle  -  \\
  - \frac{i}{2}\int\limits_{ - \infty }^\infty  {d^4 x_1 } \int\limits_{ - \infty }^\infty  {d^4 x_2 } \left\langle 0 \right|T\Psi _i^{} (t,{\bf{x}})\bar \Psi _k^{} (t_0 ,{\bf{x}}_0 ){\cal H}_I (x_1 ){\cal H}_I (x_2 )\left| 0 \right\rangle  \\
 \end{array}
\end{equation}
On the other hand, the one-electron matrix element of the
evolution operator $<\Psi^{(0)}_n| U(t,t_0)|\Psi^{(0)}_n>$ in this
approximation reads
\begin{equation}\label{add3}
\left\langle {\Psi _n^{} } \right|U(t,t_0 )\left| {\Psi _n^{} }
\right\rangle  = 1 - \frac{i}{2}\int\limits_{t_0 }^t {dt_1 }
\int\limits_{ - \infty }^\infty  {d^3 x_1 } \int\limits_{t_0 }^t
{dt_2 } \int\limits_{ - \infty }^\infty  {d^3 x_2 } \left\langle
{\Psi _n^{} } \right|T {\cal H}_I (x_1 ){\cal H}_I (x_2 )\left|
{\Psi _n^{} } \right\rangle
\end{equation}
Here as well as in Eq.(\ref{add2}) the contributions from the
terms associated with the vacuum-vacuum transition should be
removed. Thus, in order for the electronic Green function could be
reduced to the one-electron matrix elements of the evolution
operator, the integration in Eq.(\ref{add2}) must be limited to
the time intervals $t _0  \le t_1  \le t$ and $t _0  \le t_2  \le
t$. This proves that the positions of the poles of the electronic
Green function do not coincide exactly with the positions of the
poles of the Green operator, and hence do not determine exactly
the atomic energy levels. Nevertheless, for atomic hydrogen the
contribution from the virtual processes that should be removed by
the above limitation is relatively small, and this is a reason why
the standard bound-state QED provides high accuracy in calculating
the Lamb shift of the atomic energy levels. But this accuracy may
turn out to be not sufficient for the derivation of the proton
radius.

The above shows that actually there is a correction $\delta^D E$
to the radiative Lamb shift in hydrogen that is hidden in solving
the problem in a standard way. But the integral in
Eq.~(\ref{delta_Ed}) that determines this correction in the
one-loop approximation diverges at infinity, because the one-loop
self-energy operator $ \Sigma _A ({\bf{p}}',{\bf{p}};E)$ behaves
in the limit $|E| \to \infty$ as
\begin{equation}\label{sigma3}
\Sigma _A ({\bf{p}}',{\bf{p}};E)\tend \limits_{|z| \tend
\infty}\frac{\alpha }{{4\pi }}\gamma _0 E\ln \left( {\frac{{m^2  -
(E+i0)^2 }}{{m^2 }}} \right)
\end{equation}
The problem is that this divergence occurs after renormalization.

It should be noted, the above results can be trivially extended to
the description of QED corrections to energy levels of muonic
hydrogen.

\section{Nonlocality of the electromagnetic interaction}
We have shown that the standard methods based on the determination
of the atomic levels by the positions of the poles of the Green
function of the electron in the Coulomb field do not obey the
exact description of the atomic states. The dynamical shift
$\delta^D E$ which in the one-loop approximation is given by
Eq.~(\ref{delta_Ed}) describes the part of the Lamb shift that was
missing in the standard description and might be the origin of the
discrepancy between the radii derived from atomic hydrogen and
muonic hydrogen spectroscopy. At the same time, in describing this
dynamical shift we face the problem of the UV divergences that
cannot be cured by renormalization. In other words, the
renormalization theory fails in this case. This is not surprising
because in dealing with the Green operator constructed by using
the renormalization procedure we are going beyond the domain of
applicability of the standard theory of QED, which is limited to
deal only with the process that can be described in terms of the S
matrix or the Green functions. The "rug" has turned out to be too
small to hide the problem of the UV divergences that arise in the
consistent description of the bound states based on the employment
of the Green operator. Actually, in this case the natural
nonlocality of the electromagnetic interaction, which in
describing processes associated with the S matrix is hidden in the
renormalization procedures, manifests itself explicitly. This
means, that for the accurate calculations of the QED corrections
and, hence, for the accurate derivation of the proton radius from
spectroscopy of atomic hydrogen and muonic hydrogen, one has to
deduce the nonlocal interaction operator describing the
fundamental electromagnetic interaction. Keeping in mind that
taking into account the nonlocality of the electromagnetic
interaction which must result in improving the behaviour of the
self-energy operator at infinity, we can make some assumptions
about the value of the correction $\delta^D E$. Formally the order
of magnitude of this correction is given by $ \delta ^{\rm{D}} E_n
= \alpha^3 {\rm{A}}_0 \delta ^{(0)} E_n $, where $\delta^{(0)}
E_n$ coincides with the ordinary one-loop Lamb shift. The constant
${\cal A}_0$ should be determined by the form of the self-energy
operator. Thus the nonlocality of the electromagnetic interaction
must manifest itself in this constant and, hence, in the Lamb
shift of atomic energy levels.

It might seem that there is a significant arbitrariness in
choosing the form of the interaction operator. However, the fact
that the interaction operator $H_{int}(t_2,t_1)$  in the limit
$t_2  \to t_1$  should be close enough to the relevant solution of
the GDE, imposes strong limitations on its form. In addition, the
form of the interaction operator is constrained by the symmetries
of QED and the fact that in describing the processes associated
with the S matrix the GDE should lead to the same results that in
the standard theory of QED are obtained by using the
renormalization theory. A transparent example of this feature of
the GDE is given in Ref. \cite{prc}, where it has been shown that
after renormalization in the effective field theory of nuclear
forces the low energy nucleon dynamics is governed by the GDE with
a nonlocal-in-time interaction operator whose form is determined
by the above constraints up to a constant that is fixed by fitting
to the nucleon scattering data. Here we are not in position to
discuss the problem of the determination of the form of the
nonlocal interaction operator describing the fundamental
electromagnetic interaction in detail. But note, that there is a
hope the above limitations on the form of this operator will allow
one to determine it up to a constant the value of which could be
derived from the spectroscopy of atomic hydrogen and muonic
hydrogen. And just the discrepancy between the radii derived from
these measurements may provide us experimental information about
the nonlocal nature of the electromagnetic interaction.

\section{Outlook}
We have shown that the GDE provides a consistent way of solving
the bound-state problem in QED. In this way the energies and
vectors of the atomic states are determined by the positions and
residues of the Green operator defined by the inverse Fourier
transform~(\ref{u_g}) of the evolution operator. From
Eq.~(\ref{gz3}) it immediately follows that such states are
stationary. Formulae~(\ref{pole_eq2}) and~(\ref{psi}) derived in
this way obey the accurate determination of the energies and the
vectors of the atomic states. This is because they are derived as
an inevitable consequence of Eqs.~(\ref{evo}) and~(\ref{main})
that represent the basic quantum mechanical rules without entering
into the details of the interaction in the system.  As we have
shown the standard bound-state QED does not obey the rigorous
description of the atomic states, because strictly speaking the
problem of finding the poles and residues of the Green operator is
not reduced to the problem of finding the poles of the Green
function of the electron in the Coulomb field.

The discrepancy between the proton radii deduced from atomic
hydrogen spectroscopy and muonic hydrogen spectroscopy seems to be
so serious that it could force revisions of the fundamentals of
physics. However, a small part of what is allowed by the current
fundamental principles of physics has been realized in existing
theories. In fact these principles, as we have seen, allow
nonlocal-in-time interaction operators as generators of quantum
dynamics, while in the existing theories one restricts oneself to
the instantaneous interaction operators which generate the
Hamiltonian dynamics. And the UV divergences are just the cost for
this restriction. This means that local interactions are actually
inconsistent with the current physical principles. Thus in order
to realize all the possibilities allowed by the current principles
of physics, one has to take into account the nonlocality of the
fundamental interactions explicitly, and the GDE provides a
consistent way to succeed in this object. And what is important in
this connection is that Eq.~(\ref{evo}) and the GDE separated from
boundary condition~(\ref{Stend}) are simply relations that
represent the basic quantum mechanical rules. And these rules
impose such strong limitations on the physical processes that Eq.
~(\ref{main}) turns out to be the rule prescribing how to build
the probability amplitudes of physical processes from the "bricks"
associated with the contributions from the processes with
infinitesimal duration times of interaction. This rule is general
and is independent of physics with which we deal. The specific
physical content is introduced by choosing these "bricks", i.e.
specifying the form of the fundamental interaction operator. In
addition this interaction operator must obey the relativity and
other symmetries of the theory.

If the theory of QED were defined in this way, then it would be
finite and could provide the accurate calculations of atomic
energy levels and, hence, the accurate derivation of the proton
radius from the hydrogen atom and muonic atom spectroscopy.
However, additional experimental information is needed to specify
the nonlocal interaction operator. And the work of Pohl and
colleagues may provide this information, which being reanalyzed in
this way may reveal unknown features of the electromagnetic
interaction. And these features may turn out to be very
surprising.

I thank Aldo Antognini for the discussions which triggered this
study.

\end{document}